\documentclass[aps,prf,reprint,groupedaddress]{revtex4-2}

\usepackage{graphicx}
\usepackage{dcolumn}
\usepackage{bm}

\usepackage{amsmath} 
\usepackage{xspace,color}
\usepackage{lineno}
\newcommand{\mum}{$\mu$m\xspace}
\newcommand{\Ren}{\text{Re}\xspace}
\newcommand{\Win}{\text{Wi}\xspace}

\newcommand{\rb}{\mathbf{r}\xspace}

\begin{document}


\title{Canopy elastic turbulence:\\ spontaneous formation of waves in beds of slender microposts}


\author{Charlotte de Blois}
\author{Simon J. Haward} 
\author{Amy Q. Shen}
\affiliation{Micro/Bio/Nanofluidics Unit, Okinawa Institute of Science and Technology Graduate University, Onna-son, Okinawa 904-0495, Japan}


\date{\today}

\begin{abstract}
In a viscoelastic flow over a microfluidic canopy of polymeric pillars, we report the spontaneous emergence of waves in the form of propagating regions of low flow velocity compared to the surrounding flow. The occurrence of the wave is chaotic and shows characteristics of elastic turbulence. We systematically study the coupling between the low velocity wave and the microfluidic canopy by combining flow velocimetry experiments and high speed tracking of the pillars. The waves form an angle $\pm\beta$ with the primary flow direction that depends on the geometry of the pillar array. If the canopy is composed of flexible structures, the passage of a wave deflects the structures locally in a manner reminiscent of the emergence of the Monami waves observed in inertial turbulence over canopies of vegetation. Due to the analogies with classical (inertial) canopy turbulence, we name our newly-observed phenomenon as \emph{canopy elastic turbulence}.
\end{abstract}

\maketitle

\begin{figure*}[t]
\includegraphics[width=0.9\textwidth]{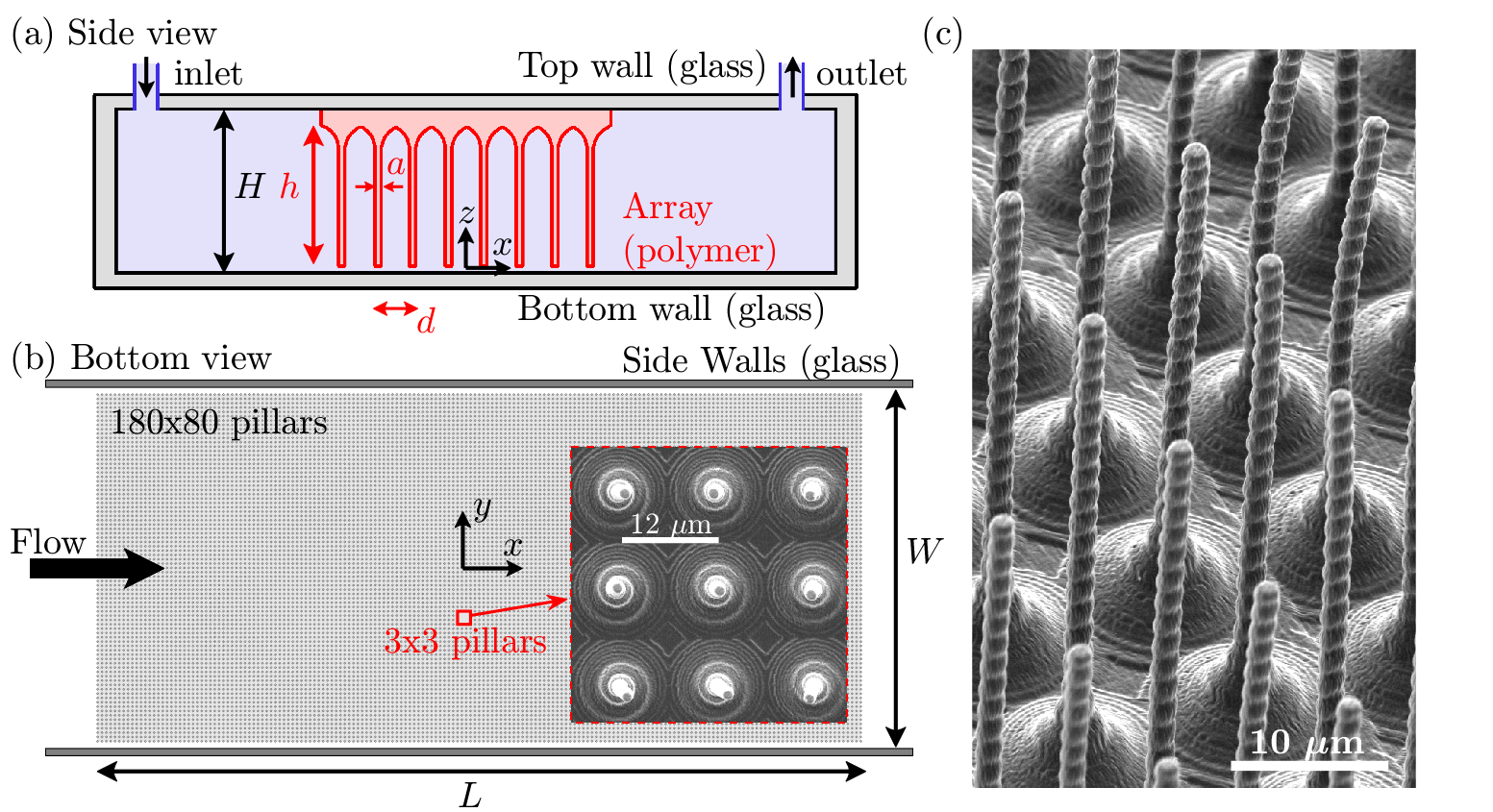}
\centering
\caption{\textbf{180$\times$80 pillar arrays in a straight microchannel.} (a) Side view ($x$,$z$) sketch (not to scale) of the microfluidic geometry. The array of polymeric pillars (in red), of height $h=48$~\mum, diameter $a=2$ \mum with uniform spacing $d=12$~\mum, is printed inside a glass microfluidic channel (in black) of height $H=60$~\mum. (b) Bottom view ($x$, $y$) sketch of a 180$\times$80 pillar array of length $L=2000$~\mum between the two side walls separated by a distance $W=1030\pm5$~\mum. The flow goes from left to right in the microchannel. The inset shows a 3$\times$3 pillar region inside the pillar array obtained by SEM imaging at magnification 2580$\times$ and $0^\circ$ tilt. (c) SEM image of zoomed-in pillars inside the pillar array obtained at magnification 2872$\times$ and $45^\circ$ tilt. The twisted microstructure of the pillar is due to a continuous offset in the orientation of the printing planes, introduced to avoid intrinsic anisotropy of the pillar that would influence its bending direction.}
\label{fig:Chip}
\end{figure*}

\section{Introduction}

Turbulence is known to appear for a Newtonian fluid when inertial stress dominates over viscous stress, i.e. at high Reynolds number ($\Ren=\rho U a /\eta$, with $\rho$ the fluid density, $U$ the flow velocity, $a$ the characteristic dimension of the system and $\eta$ the viscosity of the fluid). In contrast, for viscoelastic fluid flows, additional elastic effects can dominate the flow dynamics. When elastic stress dominates over viscous stress, i.e. for sufficiently high Weissenberg number ($\Win=\dot{\gamma} \tau$, with $\dot{\gamma}  $ the characteristic shear rate of the flow and $\tau$ the relaxation time of the fluid), viscoelastic flows can become chaotic and exhibit so-called `elastic turbulence' even at negligible inertia~\cite{Groisman2000,Pan2013,Li2012,Steinberg2021}. Although the inertial turbulence and elastic turbulence come from completely different physical origins, conceptually, both regimes are caused by the non-linearity present in the flow~\cite{Groisman2000,Browne2020a}, from the convective effect in the former case and from the nonlinear constitutive equations of the fluid in the latter case. Both affect the response time of the fluid: inertia tends to resist the changes in the flow velocity and elasticity tends to resist the deformation of the fluid. 

Many experimental observations on elastic turbulence phenomena show a similarity in features with their equivalent in the inertial turbulence regime, such as the formation and chaotic fluctuation of wakes behind obstacles~\cite{Haward2018,Haward2020,Varchanis2020}, the oscillation of flexible pillars under viscoelastic flow~\cite{Dey2018,Dey2020,Haward2019,Hopkins2020}, or the formation of eddies upstream of constrictions~\cite{Browne2020} and in porous media~\cite{Kumar2021}. Very recently the emergence of Alfvén-like waves~\cite{Varshney2019,Steinberg2021} was reported in a viscoelastic creeping flow between two obstacles positioned in a straight channel. The emergence of a Kelvin–Helmholtz-like instability~\cite{Grilli2013,Jha2021} was also observed in the same geometry~\cite{Varshney2018} with the formation of a mixing-layer (a free shear layer between flows of very different velocities unstable to small perturbations) between the two obstacles.

Canopy flow~\cite{Brunet2020} is the flow over elements of slender filaments (rigid or flexible canopies) such as plants or buildings. In the regime of inertial turbulence, a mixing-layer instability occurs in the canopy flow~\cite{Ghisalberti2002} where coflowing streams of different velocities develop above and inside the canopy, with an inflectional mean velocity profile~\cite{Finnigan2000}. The resulting mixing layer eventually becomes unstable, leading to a Kelvin–Helmholtz-like instability~\cite{Viero2017}, downstream vortex formation, and canopy turbulence~\cite{Finnigan2000}. If the canopy is composed of flexible structures such as plants, this instability induces the coherent waving of the vegetation, also referred to as Monami waves~\cite{Ackerman1993,Henry2015,Nepf2012}. 

Observing that (i) canopy turbulence in the inertial regime comes from the destabilisation of a mixing layer caused by the canopy, and (ii) mixing layer instabilities have been reported in other geometries~\cite{Varshney2018} in the regime of elastic turbulence, in this work we investigate experimentally a microfluidic canopy in the regime of elastic turbulence. We first present a new microfluidic design described in section~\ref{sec:methods} developed to study a canopy of polymeric micropillars under viscoelastic flow, with a large range of Weissenberg number ($30 < \text{Wi} < 1100$) and at negligible inertia ($\Ren < 0.1$). In section~\ref{sec:thick}, we study the case of a rigid canopy and report the spontaneous emergence of waves in a microfluidic canopy under viscoelastic flow above a critical $\Win_c\simeq120$. Such waves are not observed under a Newtonian flow in the same conditions (section~\ref{sec:Newt}). In section~\ref{sec:thin}, we switch to a flexible canopy and observe that the waves deflect the pillars when propagating. We investigate and discuss the influence of different parameters such as the rheology of the fluid or the spacing of the pillar array on the wave. Finally, we investigate the physical origin of the waves and argue that the interaction between the flow and the canopy leads to a flow instability which deflects the pillars when they are flexible, in a similar fashion to the formation of Monami waves in the classic case of canopy inertial turbulence.

\section{Materials and Methods\label{sec:methods}}

\subsection{Design and fabrication of the microfluidic devices}

In order to design a large array of polymeric micropillars inside a rectangular glass microchannel that confines the pillars, we develop a new microfluidic design shown in Figure~\ref{fig:Chip} by combining two microfabrication techniques: (i) a subtractive three-dimensional (3D) printing process called selective laser-induced etching (SLE~\cite{Gottmann2012,Burshtein2019a,Meineke2016}), for the design of an optically transparent glass channel ($H=60 \pm 5$~\mum in height, $W=1030 \pm 5$~\mum in width, and 15~mm in length) and (ii) a two-photon polymerization process for the 3D printing of a 180$\times$80 array of slender polymeric cylinders of diameter $a=4$~\mum (rigid pillars) or $a=2$~\mum (flexible pillars) and height $h=48$~\mum, spaced at $d=12$~\mum in $x$ and $y$ directions (except in section~\ref{sec:wave} where the effect of pillar spacing is investigated). The array of pillars is directly printed inside the glass. Sketches of the pillar array inside the microfluidic chip are given in Figure~\ref{fig:Chip}(a) in side view and (b) in bottom view. The square array consists of 14,400 pillars. The array length and width are $L=2$~mm and 980~\mum respectively, so that there are two empty regions of gap $25\pm 5$~\mum between the pillar array and the side walls and a space of $10\pm5$~\mum between the tip of the pillars and the bottom wall. The pillars are made of IP-dip resin (autofluorescent acrylate-based polymer, Nanoscribe GmbH), that is fluorescent under an excitation between 400 and 700~nm and have a Young's modulus of $E=2.91$~GPa. All imaging is done through the bottom wall of the device. The fabrication steps are described below.

\begin{figure*}[t]
\includegraphics[width=0.9\textwidth]{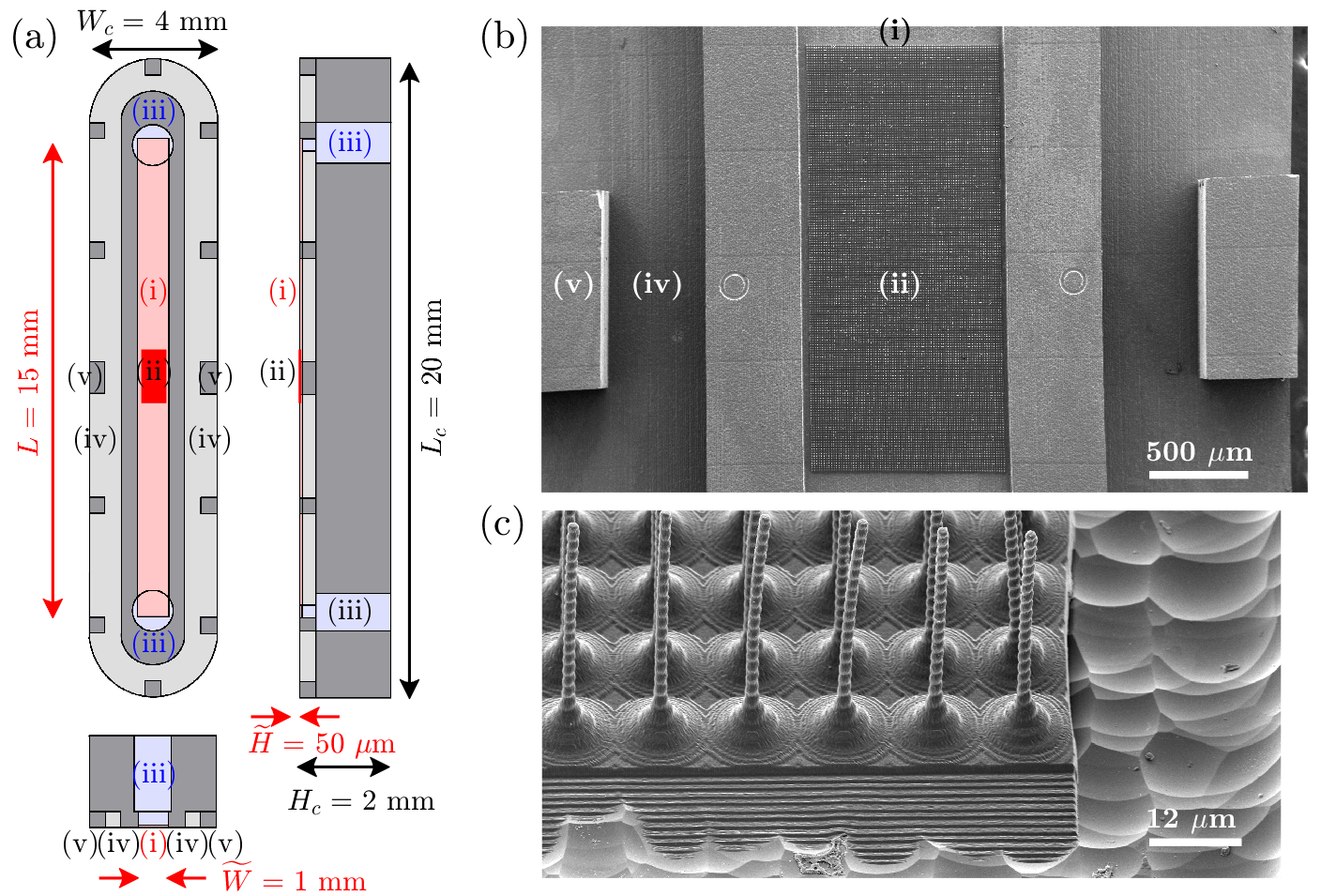}
\centering
\caption[Microfluidic chip]{\textbf{Microfluidic device:} (a) Schematic of the microfluidic device used for the experiments. (i) Main channel, (ii) arrays, (iii) inlet and outlet, (iv) gutter, (v) Pillar of support. (b) Bottom view of SEM image of the whole pillar array inside the microfluidic channel. (c) Tilted view of SEM image at an angle 45$^\circ$ of the corner edge of the pillar array.}
\label{fig:SChip}
\end{figure*}

\subsubsection{Glass channel}

An open microfluidic chip schematized in Figure~\ref{fig:SChip}(a) is designed using the 3D design software Rhinoceros\textsuperscript{\textregistered}~V5. The outer dimensions of the chip are $H_c = 2$~mm in height, $W_c= 4$~mm in width, and $L_c = 20$~mm in length. As designed, the microfluidic channel (i) has inner dimensions of $\Tilde{H} = 50$~\mum in height, $\Tilde{W} = 1$~mm in width, and $L = 15$~mm in length. The inner channel is open on the bottom part of the chip. Each end of the inner channel is connected by a hole to the upper side of the chip, which will be used to insert the inlet and outlet connections (iii). These two open holes consist of a square section of 1~$\times$~0.4~mm with 0.5~mm in height, followed by a cylindrical section of 0.63~mm in diameter and 2.45~mm in height that reaches the upper side of the chip. The cylindrical section is adapted to the size of the microfluidic connector (inlet/outlet) used, and the square section is used to separate the inlet/outlet connectors from the channel. 
On the bottom side of the chip, a gutter (iv) of 0.5~mm in depth is made around the channel and at a distance 0.5~mm from the channel. Twelve pillars (v) are created at the periphery of the chip to support it. The gutter is created to facilitate the spreading of the glue by capillarity when the open microfluidic chip is glued to a glass slide. Marks are engraved on either side of the center of the chip (the two circles seen in Figure~\ref{fig:SChip}(b)), that are used to align the pillar array of polymeric pillars with the channel.

SLE is a state-of-the-art glass etching technique to fabricate 3D glass microchannels. We follow the method developed in~\cite{Burshtein2019a}. Using the LightFab 3D printer, a 2~mm thick fused silica glass is exposed to laser radiation following the 3D design described above. The laser path defines the edges and volumes of the glass that are to be removed since exposure to the laser radiation renders the fused silica highly susceptible to chemical etching. The fused silica glass is next plunged into a KOH ultrasonic bath at 85$^{\circ}$C to etch overnight. The etched device is rinsed with deionized water. The final device has a surface resolution of typically 2-5~\mum which is observable under the microscope (see Figure~\ref{fig:SChip}(c)). For our application, the surface roughness is advantageous since it increases the surface of contact between the polymeric array and the glass chip and thus improves their adhesion.

\subsubsection{Arrays of pillars}

A base unit array is designed using the 3D design software Inventor\textsuperscript{\textregistered}, to fit the printing window of the Nanoscribe 3D laser lithography system. The unit array is composed of $8 \times 8$ cylindrical pillars linked through a continuous base of thickness 5~\mum and a fillet between the pillar and the base of radius 10~\mum is added to improve the mechanical strength of the pillars. The base improves the adhesion of the pillar array with the glass and fills the roughness of the glass chip (Figure~\ref{fig:SChip}(c)). The design is then converted to a 3D print file using the DeScribe software. The slicing distance (vertical resolution) is set at 0.3~\mum and the hatching distance (horizontal resolution) at 0.2~\mum. The hatching direction (orientation of the illumination in the plane) is set to vary by steps of 45$ ^\circ$ between each plane to avoid any anisotropy in the pillar (that leads to a preferential direction when bending). Finally, a large array is designed by encoding a spatial loop of the base array in the printing file, so that the final array has a dimension of $960 \times 2100$~\mum, which corresponds to 14,400~pillars. 

The array (ii) is printed directly inside the open microfluidic chip (described above), by two photon-polymerization of a photoresin, using the Nanoscribe. The glass chip is rinsed beforehand with isopropanol and deionized water, dried with an air gun and left 10 min on a 150$^\circ$C hot plate. A droplet of the photoresin IP-dip (Nanoscribe) is deposed inside the channel. The device is placed on the Nanoscribe platform using a 63~$\times$ objective, in dip-in liquid, lithography (Dill) mode. The laser power is fixed to 30~mW (60\% of the maximum laser power) and the scanning speed is set to 10~mm/s. 
To avoid any drift in height during the printing, the device is left for 1~h at rest on the Nanoscribe platform. Finally, the base of the pillar array is embedded in the glass surface roughness by manually finding the interface and fixing a virtual interface to be 3~\mum under the real interface. If necessary, height corrections are also added inside the printing loop if the glass interface is not perfectly flat. The total printing time for the described array is $\sim$22~h for the flexible pillars array and $\sim$44~h for the rigid pillars array.
Removal of the uncured resin while preserving the structure of the pillar array is the most challenging task of this process and limits the density at which the pillars can be printed. The smallest interpillar spacing chosen in this work ($d=12$~\mum) is the smallest distance that we are able to achieve without causing the pillars to collapse when the uncured resin is rinsed away. This was achieved by developing a three step post-printing process. The array is first developed for 1~h in 1-methoxy-2-propyl acetate (PGMEA, 484431 Sigma Aldrich), then rinsed for 1~h in isopropanol while exposed to a weak UV light (LUXO\textsuperscript{\textregistered} Ultraviolet Inspection Lamp), and finally plunged for 30~s in NOVEC\textsuperscript{TM} 7100 Engineered fluid, whose low surface tension helps to prevent the collapsing of pillars during air drying. 

Finally, the open microfluidic chip containing the micropillar array is bonded to a glass slide by using a two-step gluing process. First the chip is placed on a glass slide and a surface insensitive adhesive (Loctite 401 instant adhesive) droplet is deposited along the edge of the chip. The adhesive enters the gutter region in the microfluidic device by capillarity without entering the channel, sealing it from the outside. After drying for 20~minutes, an epoxy glue (Loctite quick mix LQM-014) is used to glue the inlet and outlet connectors and to cover the whole glass chip, to avoid any leakage and mechanically strengthen the chip. The device is ready to use after another 24~h of drying. Microscopy imaging is done through the bottom glass slide. With time and under flow, the printed array lose its adhesion to glass. Therefor one array can typically be used only for one series of experiments over one day. 

Note that because of the very long printing time of the rigid pillar array and the limited availability of the 3D printer, the use of the flexible pillar array was favored in the result section to explore the effect of different parameters such as the rheology of the fluid or the inter-pillar distance.

\subsection{Viscoelastic fluids}
\begin{figure*}[t!]
\includegraphics[width=0.85\textwidth]{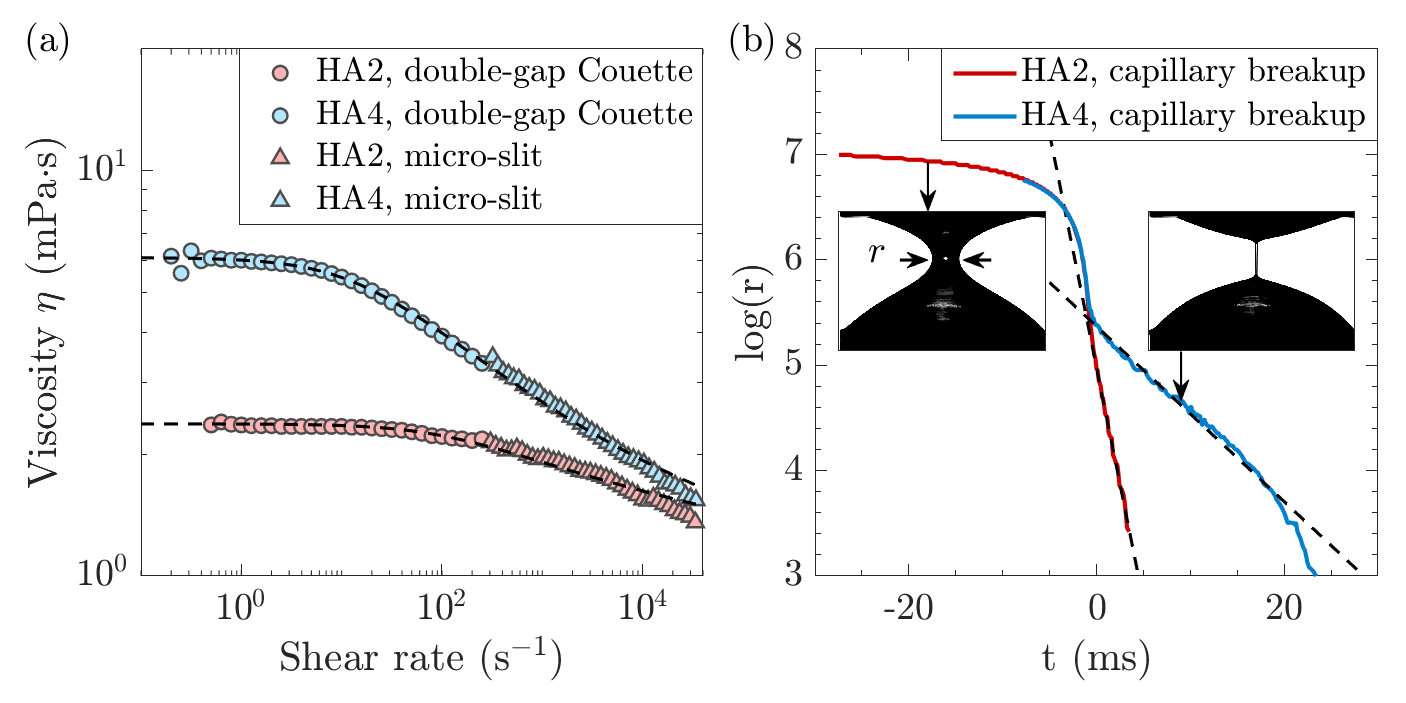}
\centering
\caption[Rheology]{\textbf{Rheology of the viscoelastic fluids:} (a) Test fluid flow curves of viscosity, $\eta$, as a function of the shear rate for solutions of 0.04~wt\% of HA2 (in red) and HA4 (in blue) in PBS. For low shear rates (circles), data are obtained by using a stress-controlled shear rheometer with a double gap Couette geometry. For high shear rates (triangles), data are obtained using a microfluidic slit rheometer. The black dashed lines represent fits to the data using the Carreau–Yasuda GNF model (equation~\ref{eq:CY}). (b) Evolution of the liquid bridge diameter with time in a Capillary Break-up Extensional Rheometer (CaBER). The dashed lines are the exponential fit of the radius in the elastocapillary regime.}
\label{fig:Rheo}
\end{figure*}

Two hyaluronic acids of different molecular weights are obtained from different sources. The sample HA2 ($M_w = 2.6 \times 10^6$~g~mol$^{-1}$) is produced by streptococcus fermentation (procured from Lifecore Biomedical, Chaska, MN). The sample HA4 ($M_w = 4.8\times 10^6$~g~mol$^{-1}$) is isolated from cockerel comb (donated by Bohus Biotech Str\"{o}mstad, Sweden). 
Viscoelastic solutions of hyaluronic acid at 0.04~wt\% are made by dissolving hyaluronic acid in a physiological phosphate-buffered saline (PBS at 0.01~M and pH~7.4, Sigma Aldrich). To avoid mechanical degradation of the polymers, special care is taken during the mixing of the solutions. The solutions are left for 48~h on a mixing roller (IKA) at 10~rpm to completely dissolve the hyaluronic acid. Samples are stored at ambient temperature and used within a week. These hyaluronic acid samples are taken from the same sources as those used in Ref~\cite{Haward2014}.

We perform shear rheometry of these fluids (Figure~\ref{fig:Rheo}(a)). The flow curves of shear viscosity at low shear rates are obtained by using a double gap Couette (Anton-Paar, DG27/T200/SS), of cup inner diameter 23~mm, outer diameter 29~mm, and bob inner diameter 23~mm, outer diameter 27~mm on a stress-controlled rotational rheometer (Anton Paar MCR 501). For high shear rates, the flow curves of viscosity are obtained using a microfluidic slit rheometer (m-VROC, RheoSense, Inc.) with a 50~\mum wide channel. The flow curves of the shear viscosity are fitted using the Carreau–Yasuda generalized Newtonian fluid (GNF) model, given by:
\begin{equation}
 \eta=\eta_\infty+(\eta_0-\eta_\infty)\Bigr[1+(\dot{\gamma}/\dot{\gamma}^*)^b\Bigl]^{(n-1)/b},
 \label{eq:CY}
\end{equation}
where $\eta_0$ is the zero-shear viscosity and $\dot\gamma ^*$ is the characteristic shear rate for the onset of shear thinning, $\eta_\infty$ is the infinite-shear-rate viscosity, taken equal to 1~mPa~s, and $b$ is a dimensionless parameter taken equal to one (in agreement with Haward et al.~\cite{Haward2014}). Repeating each experiment two times at low and high shear rates for both viscoelastic fluids, the fittings give: \mbox{$\eta_{0,\text{HA2}}=2.37\pm0.02$ mPa s}, and \mbox{$\eta_{0,\text{HA4}}=6.11\pm0.05$ mPa s}.

The relaxation times of the fluids at 25$^\circ$C are measured using the slow retraction method~\cite{Campo-Deano2010} implemented on a Capillary Break-up Extensional Rheometer (Haake CaBER 1, Thermo Scientific), with circular plates of diameter 6~mm. After deposition of the liquid between the two plates it forms a stable liquid bridge. We increase manually the distance between the plates in a quasi-static fashion, until the break-up of the bridge is observed (see inserts of Figure~\ref{fig:Rheo}(b)). The necking filament is visualized with a resolution of 8.3 \mum/pixel and the break up process is recorded at 8600 frames per second using a high speed camera. Images are thresholded and edge detection is used to find the minimum filament diameter as a function of time. The evolution of the bridge diameter with time before the break-up is presented in Figure~\ref{fig:Rheo}(b). Two successive regimes are observed, the viscocapillary regime followed by the elastocapillary regime, where the viscous and the elastic stresses oppose the thinning of the bridge respectively. The liquid bridge diameter decays exponentially with time in the elastocapillary regime~\cite{Anna2001}, with a characteristic time that is three times the relaxation time $\tau$. %
The relaxation time of the fluids is extracted by direct curve fitting of the exponential decay in this regime. The experiment is repeated 10 times for fluid HA2 and HA4, and the average relaxation times are computed with their standard deviation: $\tau_{\text{HA2}}=0.71\pm0.03$~ms and $\tau_{\text{HA4}}=3.7\pm0.2$~ms. 

For reference, we also use a Newtonian fluid of viscosity $\eta=5.2$ mPa~s that is similar to the zero-shear viscosity of the two viscoelastic fluids. The Newtonian solution is made of a mixture of Milli-Q\textsuperscript{\textregistered} water and glycerol (Sigma Aldrich) at a concentration of 50~wt\%.

\subsection{Image acquisition and processing}

The microfluidic chip is oriented so that the pillars point vertically downward. The array is observed from beneath via an inverted microscope (Nikon Eclipse Ti) with 4$\times$, 10$\times$ and 20$\times$ Nikon PlanFluor objective lenses (numerical apertures 0.13, 0.30 and 0.50 with resolutions 0.5 pix/\mum, 1.25 pix/\mum and 2.5 pix/\mum). The flow rate, controlled in the range $0.05\leq\Phi\leq 1$~mL/min by a Nemesys syringe pump (Cetoni GmbH), is imposed at the inlet of the chip. The range of imposed flow rates corresponds to a range of average flow velocity in the main microchannel of $0.02\leq \langle U \rangle =\Phi/(WH) \leq0.$4~m/s. Within this range, the largest Reynolds number reached in the pillar array is $\Ren=\rho \langle U \rangle a/\eta=0.08$. All image acquisitions are done at constant flow rates and captured 10~s after turning on the flow to avoid any transient effects. All image processing is performed in Matlab\textsuperscript{\textregistered}. 

Four parameters ($\beta$, $\alpha_w$, $\theta$, $w$) are defined to characterize the pillar displacement and the waves, illustrated in Figure~\ref{fig:coo}. The waves take an elongated shape that forms an angle $\beta$ with the $x$-axis. Inside the waves, the pillars are deflected by a distance $w$ and with an angle $\theta$ to the $x$-direction, while the flow inside the wave takes a direction $\alpha_w$. Below we describe the process used to extract these quantities. 

\subsubsection{Flow velocimetry~\label{sec:PIV}}

Micro-particle image velocimetry (micro-PIV) measurements are conducted on a Nikon Eclipse Ti inverted microscope by seeding the fluids with fluorescent particles (0.2~\mum diameter Fluoromax red, Thermo Scientific Inc.) at concentration $\simeq0.02$ wt\%. The PIV is done by cross-correlating the particle locations between pairs of successive images separated by a time 20~$\mu$s. Image pairs are acquired at an acquisition rate of 1000~Hz under a 20× objective (image resolution 2.5~pix/\mum) using a Phantom Miro high speed camera (Vision Research Inc). The measurement depth for a Nikon 20$\times$ objective lens (NA = 0.45) and 0.2~\mum particles is 11.4~\mum~\cite{Meinhart2000}. The reconstruction of the flow field is performed using the PIVlab~\cite{Thielicke2014} toolbox of Matlab\textsuperscript{\textregistered}. After subtraction of the background computed from the average of all images, the PIV processor uses cross-correlation between two successive images in four passes of respective interrogation areas of 256, 128, 64 and 32~pixels and with a window overlap of 50\%. The smallest window corresponds to the smallest distance $d =12$~\mum between two pillars and typically contains ten tracer particles. The final spatial resolution of the mapping of the flow field is then 6 \mum/pixel. The flow field is corrected by replacing the extreme values (the 5th percentile) of the velocity distribution by a local mean filter and has been averaged over five images (5~ms during which the wave propagates one third of its width) to reduce the experimental noise.

\begin{figure}[t]
\includegraphics[width=0.5\textwidth]{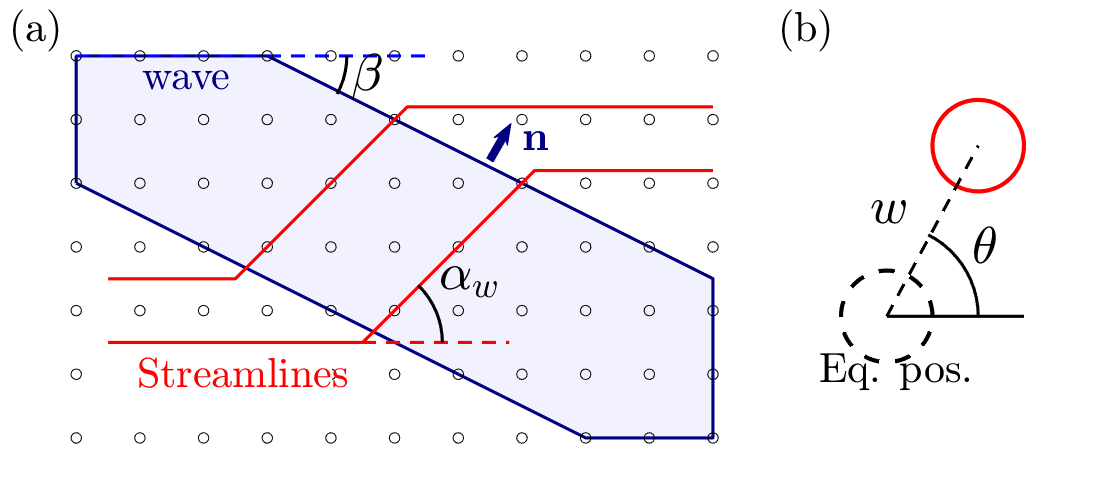}
\centering
\caption[Coordinates]{\textbf{Schematic of key parameters defined for wave characterizations:} (a) Schematic of the wave propagation, $\mathbf{n}$ the direction of propagation of the wave, $\beta$, the orientation of the wave, $\alpha_w$ the direction of the flow inside the wave. (b) Schematic of the pillar deflection, $\theta$ the angle of deflection of the pillars inside the wave, associated with the deflection $w$ of the pillar from its equilibrium position.}
\label{fig:coo}
\end{figure}

The velocities inside and outside the waves are determined by finding the peaks of the distribution of the flow velocity $U$ in time and space. The wave regions are discriminated using the first peak of the velocity distribution. The average direction of the flow $\alpha_w$  in the waves as illustrated in Figure \ref{fig:coo}(a) is computed by fitting the distribution of the flow direction inside the wave in space and time by a sum of two Gaussian distribution. 

Finally, the power spectral density of the time-resolved synchronization and flow orientation signals are computed using the pwelch function of Matlab\textsuperscript{\textregistered}.

\subsubsection{White Light images~\label{sec:WL}}

White light images are acquired using a Phantom Miro high speed camera and a Nikon Eclipse Ti inverted microscope with a 4× objective lens to visualize the whole array. A total of 3000 consecutive images are acquired at an acquisition rate of 1500~Hz. A background computed from the average of all images in time is subtracted from the images, so that the displacements inside the pillar array are highlighted, and a Gaussian filter is applied to each image. The images intensity is inverted so that the darker regions correspond to a strong fluctuation from the background image, indicating the propagation of waves. We use the observation that the average of the image intensity over one direction has the strongest signal when this direction is perpendicular to the direction of the wave to find the directions of each wave. The typical direction of the waves $\beta$ (Figure \ref{fig:coo}(a)) is taken as the average of all measured values. 

\begin{figure*}[t]
\includegraphics[width=0.95\textwidth]{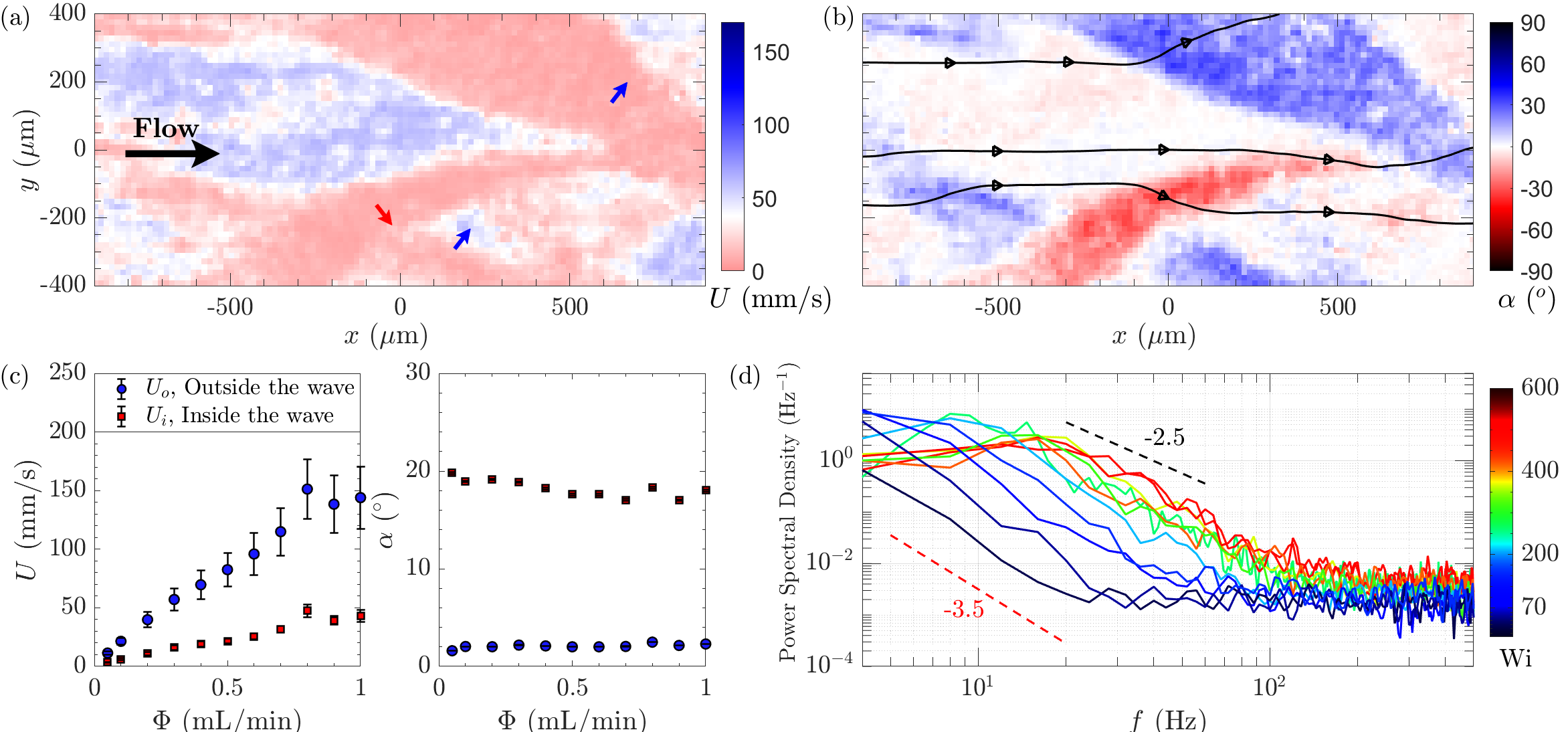}
\centering
\caption{\textbf{Viscoelastic flow over the rigid canopy ($a=4$~\mum):} Color-map of (a) the instantaneous velocity magnitude $U$ and (b)  the instantaneous flow direction $\alpha$ in the focal plane at $z=0$ for a flow rate $\Phi=0.4$~mL/min with viscoelastic fluid HA4, at a snapshot in time when a wave is observed. The black arrow in (a) indicates the direction of the flow, and the blue and red arrows the directions of propagation of the positive and negative wave, respectively. Three streamlines in (b) are shown as continuous black lines. A movie of the corresponding experiment is also given in the Supplementary~\cite{Movie1}. (c) Time and space averaged flow velocities and flow direction of a viscoelastic fluid HA4 outside the waves (blue circles) and inside the waves (red squares), measured in the focal plane at $z=0$. (d) Power spectral density of the flow direction averaged on the wave profile and at the position where the waves are the strongest, for viscoelastic fluid HA4. The colormap represents Wi numbers ranging from $30$ to $600$. The black dashed lines represent a -$2.5$ power index and the red dashed lines a -$3.5$ power index for visual comparison. }
\label{fig:Fix}
\end{figure*}

\subsubsection{Pillar tracking~\label{sec:track}}

The IP-dip resin (acrylate-based polymer, Nanoscribe) used for printing the pillar array is fluorescent under an excitation between 400 and 700~nm. Focusing on a region of interest around the center of the pillar array covering $\sim20$\% of the whole array (3000 pillars) and using a higher power (10$\times$) objective lens and laser light (wavelength 527~nm), we excite fluorescence of the pillars. A total of 3000 consecutive images obtained with the Phantom Miro high speed camera at an acquisition rate of 500~Hz are used for the tracking of the pillars. The position of each pillar and their polar coordinates ($w$, $\theta$) as presented in Figure~\ref{fig:coo}(b) are obtained from a two step process:

\begin{enumerate}
    \item Object detection and tracking: the images are cleaned using low-pass spatial filters. The connected objects present in the image are segmented by intensity threshold. The trajectories of the pillars are obtained by standard tracking methods. Between two time-steps, the positions of the pillars are associated by minimizing the distance between each pair of positions. 
	\item Polar coordinate: the equilibrium position (average position outside of a wave) of each pillar is measured. For each pillar, the polar coordinates ($w$, $\theta$) are computed. 
\end{enumerate}


\section{Results and Discussions\label{sec:Results}}

\subsection{Viscoelastic flow over the rigid canopy\label{sec:thick}}

We first design an array of thick pillars of diameter $a=4$ \mum, with a pillar height $h=48$~\mum and an inter-pillar distance $d=12$~\mum. In this case and within the range of flow rates explored ($\Phi<1$ mL/min), the deflection of the pillar is effectively negligible ($< 0.2$~\mum, beyond our microscopy resolution), hence the denomination of rigid canopy. We measure the flow field inside the rigid canopy through micro-PIV experiments as described in Section~\ref{sec:PIV}, with a $4\times$ objective to visualize the whole array. For a typical experiment with the viscoelastic fluid HA4, the magnitude of the flow velocity in the focal plane at the bottom edge of the canopy (i.e., $z=0$ plane as defined figure~\ref{fig:Chip}(a)), $U$ and its direction $\alpha$ are mapped in Figure~\ref{fig:Fix}(a, b) respectively, for a flow rate of $\Phi=0.5$~mL/min and a viscoelastic fluid (HA4). The corresponding video is given in the supplementary video~\cite{Movie1}. The Weissenberg number in the system is computed as $\Win =2 U_c \tau/a$, $a$ being the diameter of the pillar, $\tau$ the relaxation time of the fluid and $U_c=\Phi/WH$ the average flow velocity in the channel. 

The spontaneous formation of waves is observed for flow rates $\Phi\geq0.05$~mL/min ($\Win=30$) in the form of two symmetric elongated regions of distinctive low velocity (Figure~\ref{fig:Fix}(a)) and where the streamlines of the flow are effectively deflected (Figure~\ref{fig:Fix}(b)). A measure of the average flow velocity inside and outside the wave regions, in Figure~\ref{fig:Fix}(c) shows that they both scale linearly with the flow rate, but with different slope such that the difference between the velocity outside and inside the wave increases with increasing flow rate. The average flow direction outside the waves is low, the flow is globally oriented in the primary direction of the flow (the direction of the micro-channel). The average flow direction inside the wave is around $\alpha_w=20^\circ$, and does not depend of the flow rate. The waves themselves propagate at an angle $\beta \simeq \pm 30^\circ$ to the primary flow direction. 

We stress that these waves appear in the absence of individual activation of the pillars, purely from the interaction between the viscoelastic flow and the canopy. No wave is observed in the flow upstream or downstream the canopy. The Reynolds numbers in our system are sufficiently small ($\Ren < 0.1$) to safely rule out any effect of inertial turbulence, which is corroborated by the absence of any wave under a Newtonian flow in the same geometry (more details in section~\ref{sec:Newt}). In Figure~\ref{fig:Fix}(d) we compute the power spectral densities of the flow direction $\alpha$, for different flow rates. Although they present a peak at low frequency, the average frequency of generation of the waves which increases with increasing flow rate, the spectra exhibit a strong power-law decay at higher frequencies. The power-law slope is found to be around $-3.0$ (between $-2.5$ and $-3.5$), which is a signature of elastic turbulence~\cite{Steinberg2019a} and confirms that the waves observed in our system occur in the elastic turbulence regime. 

The limited availability of the 3D printer and the very long printing time of the array of pillars of diameter $a=4$~\mum make this array technically challenging to print in large quantity. In the following, the flexible pillar array is favored to conduct more systematic studies of the system. In this case, the pillars of diameter $a=2$ \mum are thin enough to be deflected under Newtonian or viscoelastic flow, but the deflection of the pillars remains small compared to the pillar height ($w/h < 0.06$) or spacing ($w/d < 0.25$), so that it does not significantly affect the flow field (which is confirmed by experimental observation).

\subsection{Newtonian flow over the flexible canopy\label{sec:Newt}}

\begin{figure}[t]
\includegraphics[width=0.5\textwidth]{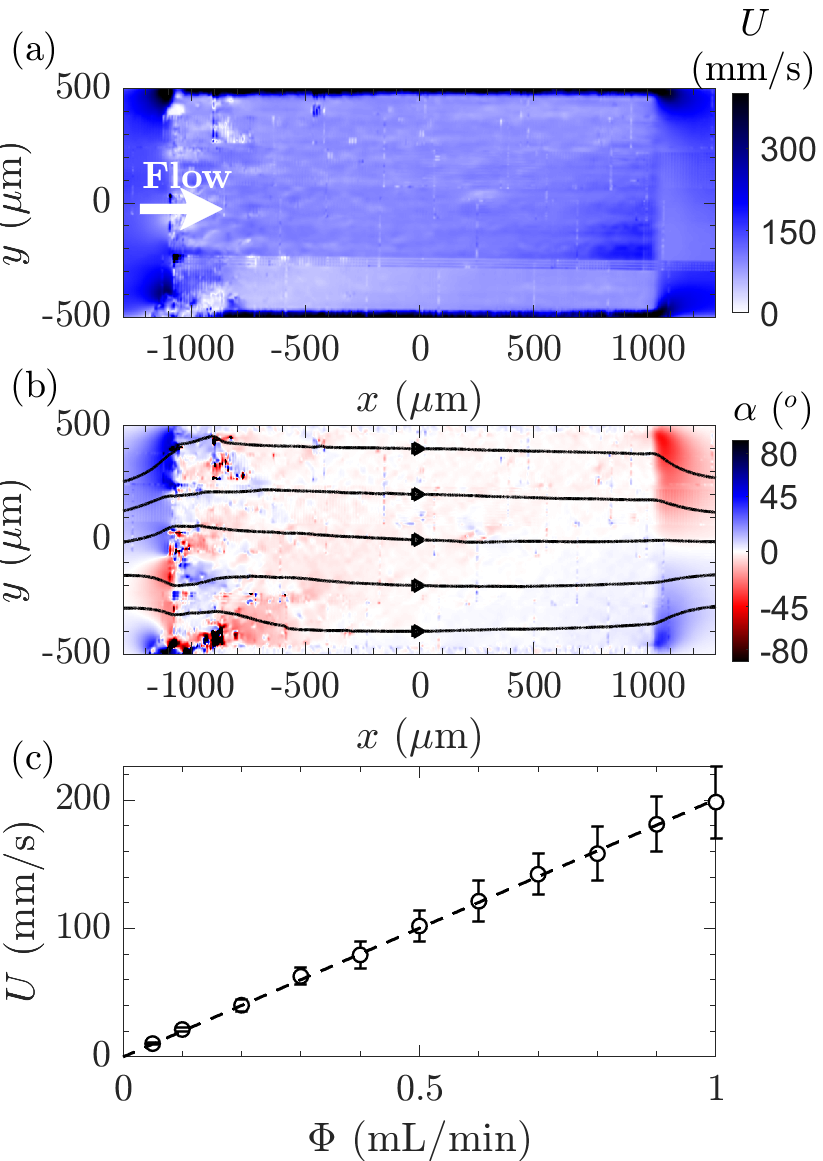}
\centering
\caption{\textbf{Newtonian flow over the flexible canopy:} (a) Color-map of the velocity amplitude $U$ and (b) color-map of the flow direction $\alpha$ reconstructed for the whole array for a flow rate $\Phi=0.4$~mL/min under a Newtonian flow. The white arrow in (a) indicates the direction of the imposed flow. Streamlines are shown in continuous black lines in (b). (c) Average flow velocity in the center region of the array.}
\label{fig:Newt}
\end{figure}

We design an array of thin pillars of diameter $a=2$ \mum, with the same pillar height $h=48$~\mum and an inter-pillar distance $d=12$~\mum as the first array. As the deflection of a pillar under a force scales as $(h/a)^4$, halving the pillar thickness increases its deflection by a factor $16$, which is observable with our microscope ($w=1-5$~\mum), hence the denomination of flexible canopy. 

Reference experiments using a Newtonian fluid (water/glycerol, 50wt\%) at flow rates $0.1\leq\Phi\leq 1$~mL/min, under white light as described in Section~\ref{sec:WL} show that the pillars are deflected by a few micrometers from their initial rest positions in the direction of the flow. The pillar displacement is steady with little variation in time and remains homogeneous in the center region of the pillar array. In all these reference experiments, no wave is observed. 

Reference micro-PIV experiments as described in Section~\ref{sec:PIV} are conducted in the pillar array under Newtonian flow, with the focal plane at $z=0$. The flow field is averaged in time over 100 images acquired at a frequency of 1000~Hz to obtain the typical flow field in a $500 \times 300$~$\mu \text{m}^2$ regions of the pillar array. The experiment is repeated to reconstruct the flow field in the whole array, as presented in Figure~\ref{fig:Newt}(a) and (b). 
We observe that a significant part of the flow goes around the pillar array in the space between the pillar array and the channel side walls (of $25 \pm 5$~\mum), which is slightly larger than the distance between two pillars ($d=12$~\mum) or between the canopy and the bottom glass surface ($\approx 10$~\mum) This affects the boundaries of the pillar array, but not the center region. In the center region of the pillar array, the Newtonian flow field is parallel along the channel direction and we observe very little variation in time and space. Finally, we measure the average flow velocity in the center region of the array where the flow is parallel, which scales linearly with the flow rate in Figure~\ref{fig:Newt}(c). 

\begin{figure*}[t]
\includegraphics[width=0.90\textwidth]{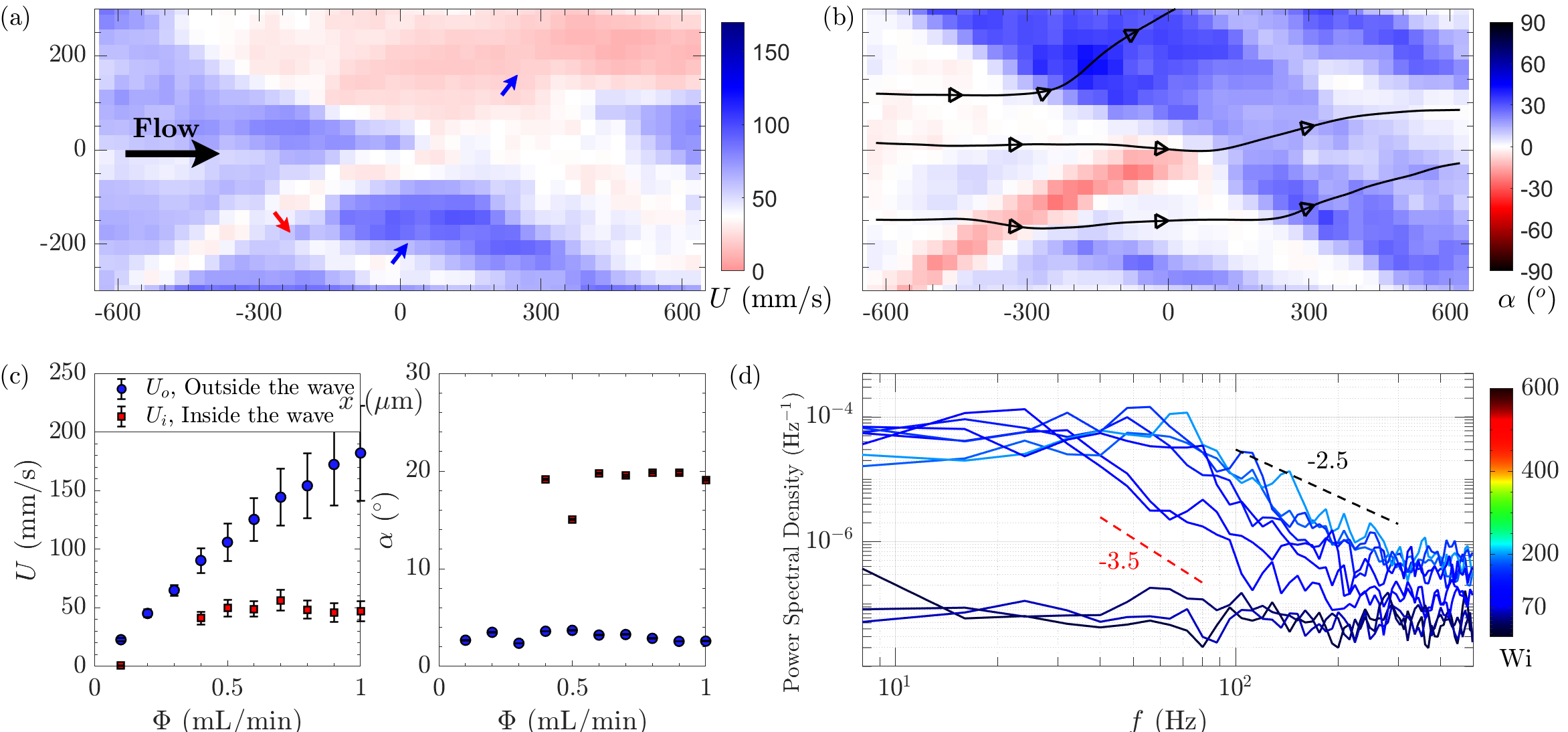}
\centering
\caption{\textbf{Viscoelastic flow over the flexible canopy ($a=2$~\mum):} Color-map of (a) the instantaneous velocity magnitude $U$ and (b)  the instantaneous flow direction $\alpha$ in the focal plane at $z=0$ for a flow rate $\Phi=0.4$~mL/min with viscoelastic fluid HA2, at a snapshot in time when a wave is observed. The black arrow in (a) indicates the direction of the flow, and the blue and red arrows the directions of propagation of the positive and negative wave, respectively. Three streamlines in (b) are shown as continuous black lines. (c) Time and space averaged flow velocities and flow direction of a viscoelastic fluid HA2 outside the waves (blue circles) and inside the waves (red squares), measured in the focal plane at $z=0$. (d) Power spectral density of the flow direction averaged on the wave profile and at the position where the waves are the strongest, for viscoelastic fluid HA2. The colormap represents Wi numbers ranging from $30$ to $600$. The black dashed lines represent a -$2.5$ power index and the red dashed lines a -$3.5$ power index for visual comparison.}
\label{fig:PIV}
\end{figure*}
\begin{figure}[t]
\includegraphics[width=0.45\textwidth]{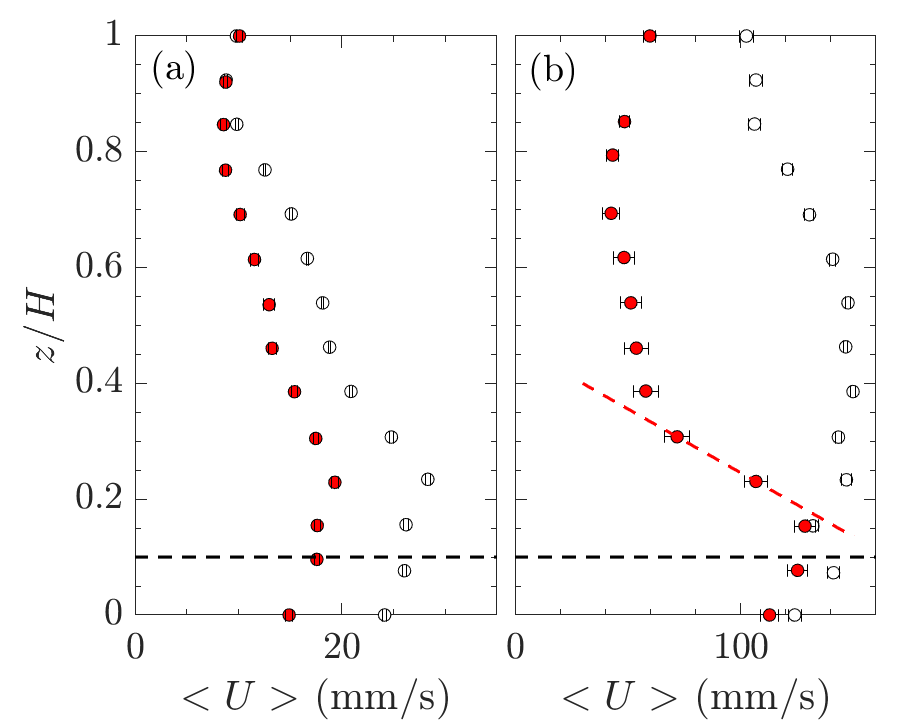}
\centering
\caption{\textbf{Vertical profiles of the amplitude of the flow} averaged in time and space for a Newtonian flow (white circles) and a viscoelastic flow (red circles), at flow rates of (a) $\Phi=0.1$~mL/min $<\Phi_c$ and (b) $\Phi=0.7$~mL/min $>\Phi_c$. At rest the pillars extend from $z/H=1$ to $z/H=0.1$ (dashed black line). The red dashed line highlight the strong variation of the velocity in $z$ when in the elastic turbulent regime.}
\label{fig:z}
\end{figure}

\subsection{Viscoelastic flow over the flexible canopy\label{sec:thin}}

\subsubsection{Flow field in the pillar array}

In the flexible canopy, micro-PIV experiments as described in Section~\ref{sec:PIV} are conducted in different $z$ planes perpendicular to the pillars. The experiments are performed with viscoelastic fluids (HA2 and HA4). The magnitude of the flow velocity in the focal plane at $z=0$, $U$ and its direction $\alpha$ are mapped in Figure~\ref{fig:PIV}(a, b) respectively for a flow rate $\Phi=0.4$~mL/min and the fluid HA2.
Similarly to the case of the rigid canopy, waves spontaneously emerge in the array in the form of low flow velocity waves that propagate at an angle $\beta\simeq 30^\circ$ to the primary flow direction. A notable difference with the rigid array is that the waves appear for a higher Weissenberg number $\Win_c\simeq120$. This effect of the object flexibility on the critical Weissenberg number of the instability is also found in~\cite{Dey2018}, for the onset of the fluctuation of a single pillar and its wake under viscoelastic flow. For $\Win>\Win_c$, the high similarity between the flow field in the rigid and flexible canopies confirms that the deflection of the pillar in these experiments is too small to significantly affect the physics at play. 

The characteristic velocities outside ($U_o$) the waves is measured for flow rates $0.1\leq\Phi\leq 1$~mL/min in Figure~\ref{fig:PIV}(c), and increase linearly with the flow rate. The flow velocity inside the wave is also measured when there are waves, and is found to be constant and independent of the flow rate. In a similar way to the rigid array case, the flow is parallel to the primary flow direction outside the waves, and is deflected at an angle $\alpha_w\simeq 20^\circ$ inside the waves.  

The power spectral densities of the flow direction for different flow rates are also measured, and the slopes of the curve at high frequencies and above $\Win_c$ are found between -2.5 and -3.5. On the contrary, for $\Win<\Win_c$ no such power law is observed. This confirms that elastic turbulence is necessary for the apparition of the waves. 

\begin{figure*}[t]
\includegraphics[width=0.95\textwidth]{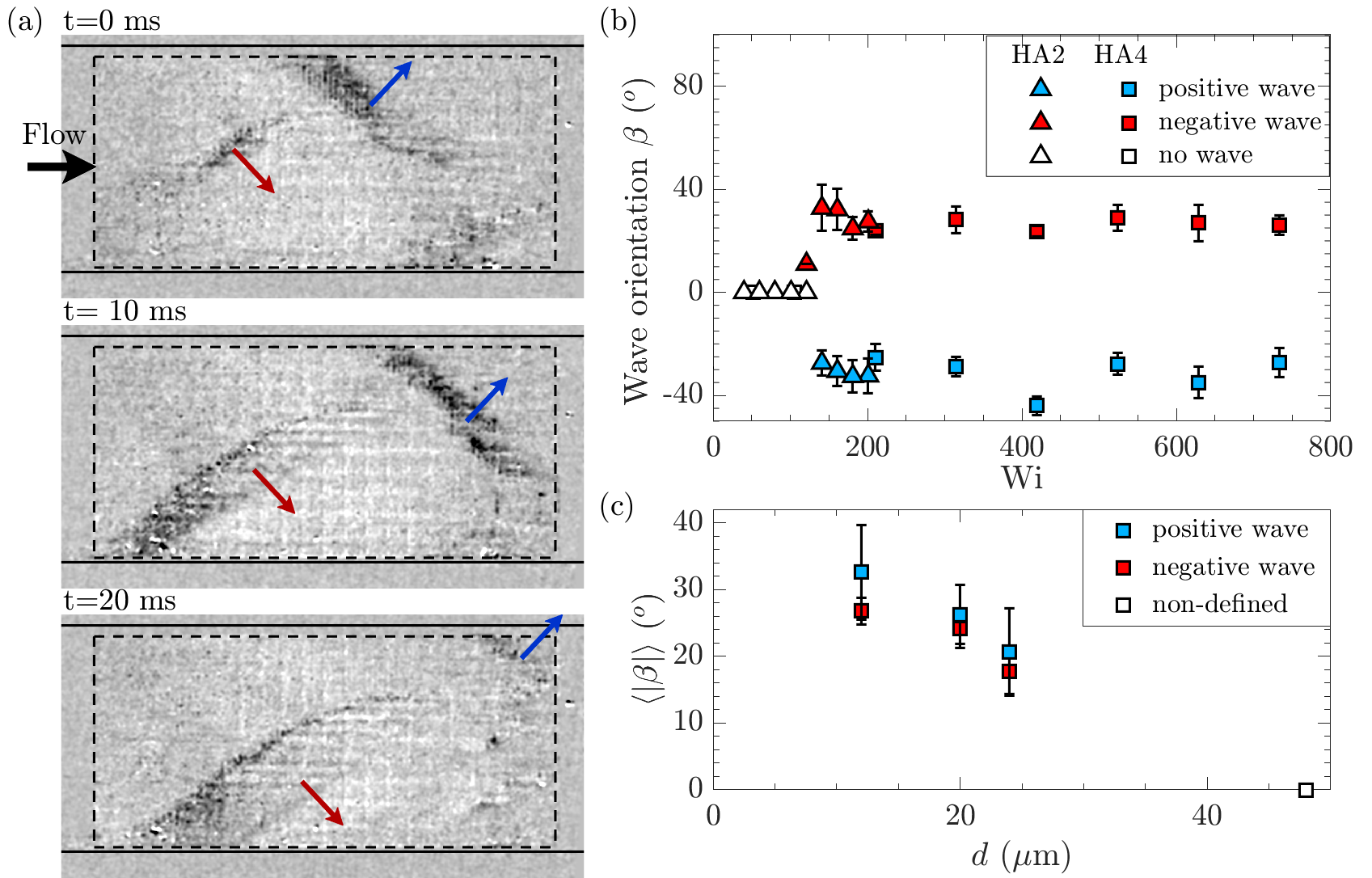}
\centering
\caption{\textbf{Spontaneous waves.} (a) Three snapshots with a 10~ms interval of an array of flexible structures in a viscoelastic flow (HA2, at a flow rate $\Phi=0.9$~mL/min) under white light. The video of the corresponding experiment is given in the Supplementary~\cite{Movie2}. The flow goes from left to right. The side walls of the channels are highlighted with black solid lines. The borders of the pillar array are highlighted with black dashed lines. The blue (respectively red) arrows indicate the direction of propagation of the positive (respectively negative) waves.  (b) Measure of the positive (blue) and negative (red) wave angles $\beta$ from the images obtained in white light, as a function of the flow rate $\Phi$ for viscoelastic fluids HA2 (triangles) and HA4 (squares). The error bars correspond to the standard deviation of the measurements over all waves. The black dashed lines mark the average angle of the waves over all experiments $\beta=\pm29^\circ \pm5^\circ$. (c) Average over all flow rate $\Phi>\Phi_c$ of the absolute positive (blue) and negative (red) wave angles $\langle \beta \rangle$ for arrays with different distance between pillars $d = 12$, 20, 24 or 48~\mum. For $d = 48$~\mum, the waves do not have a well defined orientation. The error bars correspond to the standard deviation of the angles $\beta$ measured at the different flow rates.}
\label{fig:WL}
\end{figure*}

Finally, the average flow velocity in time and space is computed in different $z$-planes between $z=0$ (bottom wall) and $z=H$ (top wall, base of the canopy) to build the vertical flow profile inside the pillar array. The results are presented in Figure~\ref{fig:PIV}(d) and (e), for Newtonian and viscoelastic flows, with two flow rates (d) $\Phi=0.1$~mL/min and (e) $\Phi=0.7$~mL/min, below and above the onset of the instability. The average velocity is smaller in the viscoelastic case, which we attribute to more fluid contouring the pillar array and flowing through the small space between the side wall and the pillars, avoiding the canopy due to the shear thinning properties of the viscoelastic fluid. For $\Win >\Win_c$, the vertical flow profile exhibits a strong variation of velocity in $z$ in the mixing layer of the canopy (around $z/H=0.3$) suggesting the presence of an inflection point. Note that the amount of data at the inflexion point is limited by the resolution in $z$ of our current set-up and dedicated systematic experiments will be needed in the future to further characterize the mixing layer of the canopy. In canopy inertial turbulence, the presence of an inflection point in a mixing layer of a canopy makes the flow susceptible to the Kelvin-Helmholtz instability (Rayleigh’s Inflection Point Criterion)~\cite{Kundu2002}.

\begin{figure*}[t]
\includegraphics[width=0.95\textwidth]{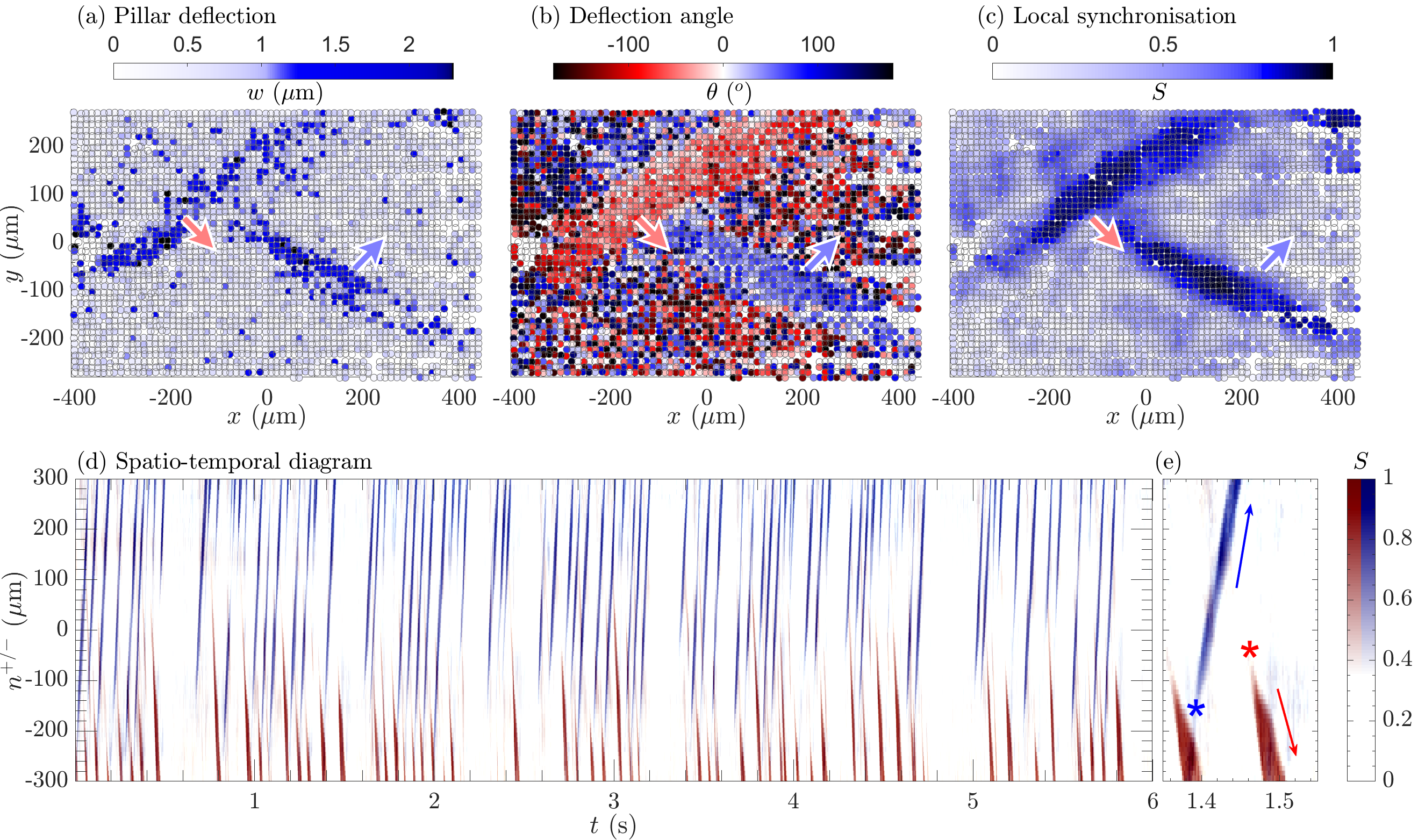}
\centering
\caption[Geometry]{\textbf{Characteristics of the waves.} Observation of a viscoelastic flow (HA2) passing the micropillar array at flow rate $\Phi=0.6$~mL/min. (a) to (c) Individual positions of pillars at a given time $t$, with colors corresponding to (a) the displacement $w$ of the pillars, (b) the deflection angle $\theta$, and (c) a measure of the local alignment of the pillars $S$ (see definition in the text). The arrows show the direction in which the waves propagate through the pillar array. (d) Superposition of the spatio-temporal diagrams of the positive wave (blue RGB channel) and the negative wave (red RGB channel). $n^{+/-}$ is a coordinate in the direction of wave displacement. (e) Zoomed-in spatio-temporal diagram between $t=1.35$~s and $t=1.55$~s. The red and blue stars indicate the time and position of the onset of the wave; the red and blue arrows display their respective wave propagation direction.}
\label{fig:Wave}
\end{figure*}

\subsubsection{Observation of the pillar deflection\label{sec:wave}}

To study the interactions between the pillars and the flow, we visualise the entire array under a white light as described in Section~\ref{sec:WL}. Viscoelastic flow experiments are conducted using the two hyaluronic acid solutions HA2 and HA4. Three snapshots of the pillar array under a white light with 10~ms intervals with a constant flow rate $\Phi=0.9$~mL/min of HA2 are shown in Figure~\ref{fig:WL}(a), and a video of the corresponding experiment is given in the Supplementary~\cite{Movie2}. 

Regions where pillars have increased deflection (darker regions in the images in Figure~\ref{fig:WL}(a)) propagate through the pillar array obliquely to the primary flow direction, which correspond to the passage of waves. In the following, we denote the waves in which the pillars deflect with $\theta>0$ as being positive, and those in which $\theta<0$ as being negative. We stress that while the synchronized fluctuating motion of the pillars enables easy visualization of the waves, it is not a requirement for the formation and propagation of the wave, and that the waves are really regions of low flow velocity compared with the surrounding, that propagate over the canopy. Given that the pillar deflection does not affect significantly the wave behavior, in the following we use the measurement of the pillar deflection to characterise the temporal and geometrical characteristics of the waves. 

The experiment is repeated for different flow rates and the angles of the waves are measured and reported in Figure~\ref{fig:WL}(b). Waves are detected above a critical flow rate $\Phi=0.6$~mL/min for HA2 and $\Phi=0.2$~mL/min for HA4, which for both correspond to a critical Weissenberg number $\Win_c \approx 120$. The waves form an angle $\beta = \pm 29^{\circ}\pm 5^{\circ}$ along the $x$ direction, which is nearly constant for all flow rates considered with both HA2 and HA4 solutions. 

The series of experiments is reproduced in different arrays with different distance between the pillars $d=12$, 20, 24 and 48~\mum, with the HA4 fluid. The emergence of waves appear at the same critical $\Win_c \approx 120$, and we measure the absolute value of the wave angle averaged over all flow rates $\Phi>\Phi_c$, reported in Figure~\ref{fig:WL}(c). For $d=12$, 20 and 24~\mum, $\langle|\beta|\rangle$ decreases with increasing $d$, while for $d=48$~\mum the propagating region does not take an elongated shape, and thus does not possess a characteristic angle. 

\begin{figure*}[t]
\includegraphics[width=0.9\textwidth]{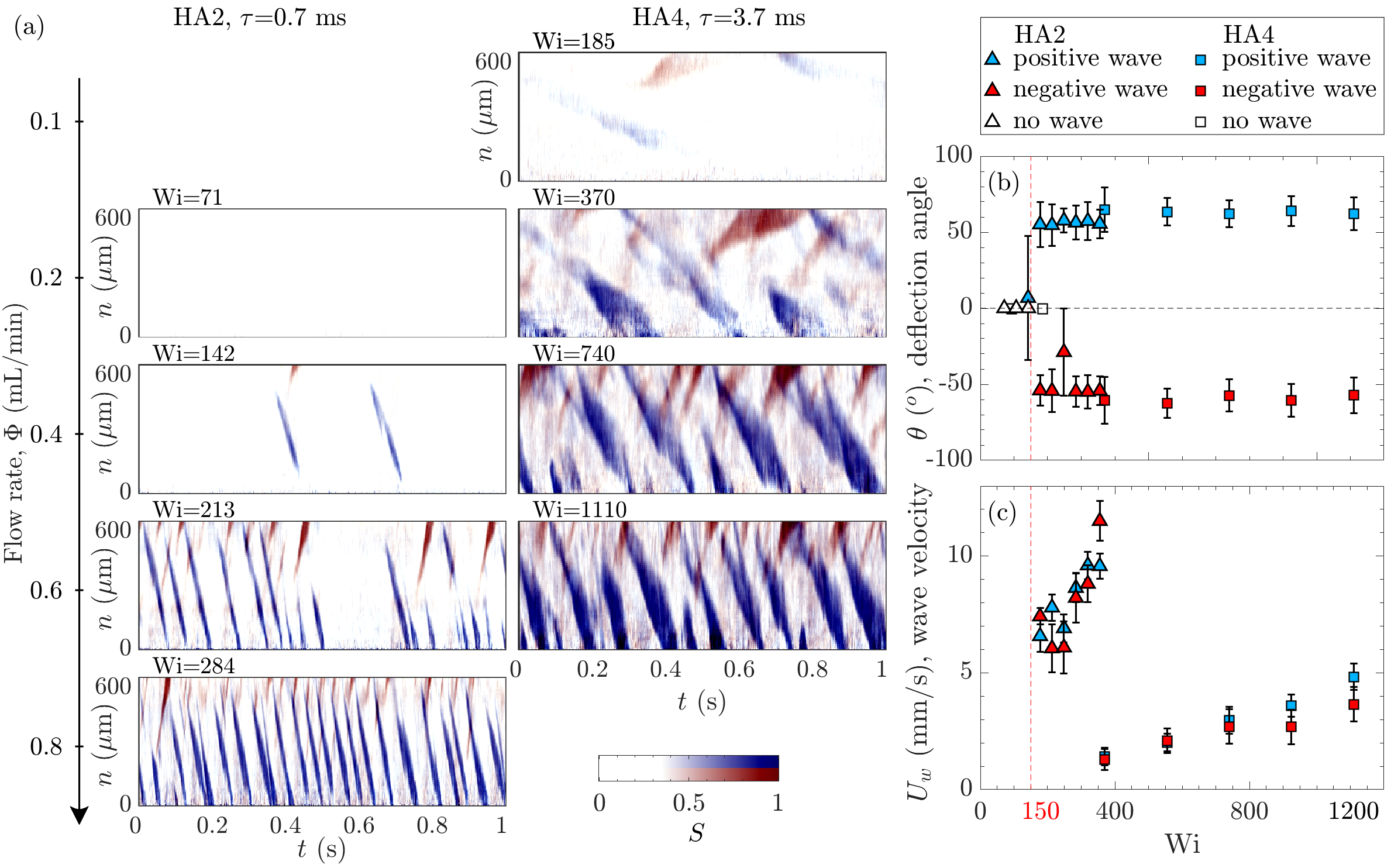}
\centering
\caption[Geometry]{\textbf{Wave evolution with the flow rate.} (a) Evolution of the spatio-temporal diagram of the waves for the viscoelastic fluids HA2 (left) and HA4 (right), for flow rates $0.1\leq\Phi\leq 0.8$~mL/min (increasing from top to bottom), with corresponding Wi numbers. The colormap corresponds to the local synchronization $S$ averaged in the direction of propagation of the wave. (b) Average orientation of the pillars inside a wave with increasing Wi for the positive (blue) and negative (red) waves. (c) Average velocity of the wave computed from the slope of the waves in the spatio-temporal diagrams, plotted against Wi. All errors bars correspond to the standard deviation of the measures for all pillars or waves. The vertical red dashed line corresponds to the onset of the wave formation at critical $\Win_c \approx 120$ for both HA2 and HA4.}
\label{fig:Flo}
\end{figure*}

\subsubsection{Tracking of the pillars deflection}

Pillar tracking experiments are conducted as described in Section~\ref{sec:track}, and focusing on a region of interest (of $\sim 3000$~pillars). Figure~\ref{fig:Wave} illustrates a typical viscoelastic flow experiment with HA2 at a flow rate of $\Phi=0.6$~mL/min. The scatter plots in Figure~\ref{fig:Wave}(a)-(c) are computed from the pillar positions recorded when both a positive wave and a negative wave were observed during the same snapshot in time. Figure~\ref{fig:Wave}(a) shows the deflection of each pillar $w$. Inside the wave, the pillar deflection
is around 2 \mum larger than it is outside the wave. Figure~\ref{fig:Wave}(b) shows that inside the wave, the pillars are uniformly deflected in the direction of wave propagation. Finally, we introduce an order parameter $S(i,t)$ that describes the local synchronization of the pillars, computed from the first moment of the orientation distribution $m_1$ in a circular region of radius $3\times d$ (i.e., encompassing 36 pillars):

\begin{linenomath}
\begin{align}
&m_1=\frac{1}{N}\sum_k^N e^{i\theta(\rb_{jk}<3d,t)},\\
&S(j,t)=\vert m_1 \vert,\label{eq:S}
\end{align}
\end{linenomath}

where the subscript ‘$j$’ is the index of a given pillar, $t$ is the time and $r_{jk}$ is the distance between pillars $j$ and $k$. Here, $S=0$ (respectively $S=1$) corresponds to a region where the pillars are randomly deflected (respectively synchronized).	

Figure~\ref{fig:Wave}(c) shows the order parameter $S$ (equation~\ref{eq:S}), which indicates that inside the wave the pillars are highly synchronized with their neighbors.

\subsubsection{Spatio-temporal diagram}

To represent the evolution of the positive and negative waves with time, we construct a spatio-temporal diagram of the local synchronization averaged over the wave profile. The local synchronization of the pillars $S$ shown in Figure~\ref{fig:Wave}(c) is averaged over the direction of the positive wave $\beta^+ = -29^\circ$ (resp. negative wave $\beta^- = 29^\circ$), to obtain the wave profile $S(n^{+/-},t)$ along its direction of propagation $n^+$ (resp. $n^-$):
\begin{linenomath}
\begin{align}
    n^+&=C_1 - x \sin(\beta^+) - y \cos(\beta^+),\label{eq:n1}\\
    n^-&=C_2 - x \sin(\beta^-) + y \cos(\beta^-).\label{eq:n2}
\end{align}
\end{linenomath}
Here $C_1$ and $C_2$ are offsets introduced for the superposition of the two spatio-temporal diagrams. For both waves $n^{+/-}$ is oriented such that low $n^+$ (low $n^-$) correspond to low $y$ (resp high $y$), so that the positive wave (in blue) propagates toward high $n^+$ and the negative wave (in red) propagates toward low $n^-$.
	
The spatio-temporal diagrams corresponding to positive and negative waves are constructed from the evolution of the respective wave's profile with time. The two spatio-temporal diagrams are finally superposed in one RGB image using the blue channel for the positive wave and the red channel for the negative wave (as shown in Figure~\ref{fig:Wave}(d)). Periods of regular wave production are interspersed by short time intervals without waves. Some zoomed-in waves are presented in Figure~\ref{fig:Wave}(e), where the initial appearance of the waves is highlighted with a star, followed by its propagation in the $n^+$ (in blue) or $n^-$ (in red) directions. 

More spatio-temporal diagrams of the local synchronization in the pillar array are shown in Figure~\ref{fig:Flo}(a) for two different viscoelastic fluids (left column for HA2 and right column for HA4) with four representative flow rates (increased from top to bottom).
At low flow rates (or low Wi) no waves are observed: the pillars are deflected uniformly and steadily under flow, as they are in a Newtonian flow. Above a critical flow rate, i.e., $\Phi > 0.4$~mL/min ($\Win = 142$) for HA2 and $\Phi >0.1$~mL/min ($\Win = 185$) for HA4, two waves with opposite angles appear. The critical Wi at which this transition occurs is similar for both viscoelastic fluids, $\Win_c\simeq 120$. With increasing flow rate (or Wi), the frequency of apparition of the wave increases. 

The synchronization parameter $S$ is used to accurately differentiate the pillars inside a wave from the pillars outside a wave. We chose a criterion $S>0.9$ to discriminate pillars in the center of a wave from the observation of the spatio-temporal diagrams. The typical deflection angles inside waves shown in Figure~\ref{fig:Flo}(b) are measured by fitting the distribution of the deflection angles $\theta$ of all pillars inside a wave by the sum of two Gaussian distributions, and extracting their mean values. The error bars correspond to the standard deviations of the Gaussian distributions. Figure~\ref{fig:Flo}(b) shows the evolution of the pillar deflection angle with Wi for HA2 (triangles) and HA4 (squares), for the positive (blue) and negative (red) waves. The angles of deflection, $\theta=\pm 60^{\circ}\pm 5^{\circ}$, do not depend significantly on Wi. We extract the velocity of the waves $U_w$ by image segmentation. Note that $\theta\simeq 90^\circ-\beta$, thus the pillars are trivially deflected in the direction of propagation of the waves. 

The measure of the waves velocity allows us to distinguish these waves from the elastic Alfven waves that come from the propagation of elastic stress in a polymer solution~\cite{Steinberg2021}, observed in a viscoelastic creeping flow between two obstacles positioned in a straight channel, in the regime of elastic turbulence~\cite{Varshney2019}. These elastic Alfven waves possess two properties that are distinctly different from the spontaneous waves observed in our system. First the elastic Alfven waves propagation speed is larger than the flow velocity, while the speed of the waves reported in this work $U_w$ is smaller than the measured flow velocity (i.e., $U_w \in 1-10$~mm/s, $U_i \simeq 50$~mm/s, $U_o \in 50 - 200$~mm/s). Secondly the elastic Alfven wave speed displayed a nonlinear dependence on the Weissenberg number, while in our system the wave velocity scales linearly with the Weissenberg number (see Figure~\ref{fig:Flo}(d)). 

\vspace{10pt}
\begin{figure}[t]
\includegraphics[width=0.45\textwidth]{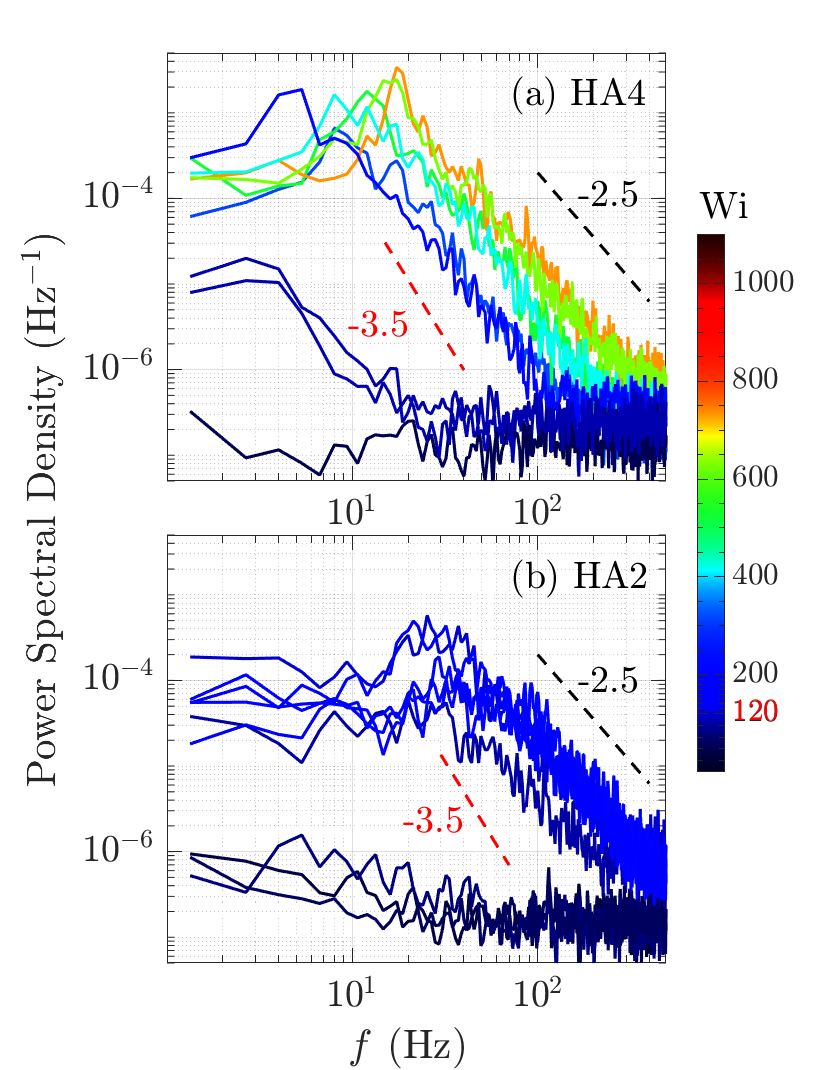}
\centering
\caption{\textbf{Elastic turbulence of Viscoelastic flow over the flexible canopy:} Power spectral density of the local synchronization from the spatio-temporal diagram at the position where the waves are the strongest, for viscoelastic fluids (a) HA4 and (b) HA2. The colormap represents Wi numbers ranging from $20$ to $1100$. The black dashed lines represent a -$2.5$ power index and the red dashed lines a -$3.5$ power index for visual comparison.}
\label{fig:Turb}
\end{figure}

\subsubsection{Power spectral densities} 

In Figure~\ref{fig:Turb} we compute the power spectral densities of the local synchronization $S$ for HA2 and HA4. Again they present a peak at low frequency, which is the average frequency of generation of the waves, and for $\Win > \Win_c$ the spectra exhibit a strong power-law decay for frequencies $f > 20$ Hz. The power-law slope is found to be around $-3.0$ (between $-2.5$ and $-3.5$), which corroborate with our previous observations that the waves occur in the elastic turbulence regime. 

\section{Conclusions and perspectives\label{sec:discussion}}

In this work, we report the spontaneous emergence of waves in a canopy under viscoelastic flow in the regime of elastic turbulence. Above a critical Weissenberg number ($\Win_c \approx 30$ for a rigid array and $\Win_c \approx 120$ for a flexible array), two symmetric elongated regions of low flow velocity compared with the surrounding, and where the flow is locally deviated in the direction of propagation of the wave propagate over the canopy. If the slender micropillars are flexible, the wave also deflects the pillars in its direction of propagation. The waves propagate with a characteristic angle $\pm \beta$ which depends on the geometry of the canopy as it decreases with increasing pillar spacing $d$.  

Although more dedicated experiments are necessary for its characterisation, the mean $z$-flow profile measured in the wave regime strongly resembles a mixing layer with the presence of an inflexion point, which in the inertial regime is considered a signature of canopy turbulence. Considering all previous observations, we suggest that low-velocity wave observed in our system are the result of an elastic Kelvin–Helmholtz-like instability, and the resulting deflection waves propagating through the structures of our flexible canopies are the elastic equivalent of the Monami waves observed in the inertial regime. By analogy, we propose the new flow phenomenon observed in our study to be termed as canopy elastic turbulence.

To control the canopy flow processes, one needs to understand the coupling and feedback of flow transport inside and outside canopies and the corresponding fluid-structure interactions. The existing research for inertial canopy flows suggests that the following key parameters are of great importance: canopy density and porosity, level of submersion and the stiffness of the individual canopy elements. The microfluidic device and the methodology we developed can be used as a model system to explore the effect of other parameters in future studies. 

Finally, a possible future theoretical work on canopy elastic turbulence could take inspiration from methods currently used to study canopy inertial turbulence, such as the two domains approach~\cite{Ochoa-Tapia1995} that model the porous medium and the fluid separately and connect them via boundary conditions. 

\subsection{Acknowledgment}

We gratefully acknowledge the support of the Okinawa Institute of Science and Technology Graduate University (OIST) with subsidy funding from the Cabinet Office, Government of Japan. Charlotte de Blois, Simon J.\ Haward, and Amy Q.\ Shen also acknowledge funding from the Japan Society for the Promotion of Science (JSPS, Grant Nos. 20K14656 and 21K03884 and 20K22403) and the Joint Research Projects (JRPs) supported by the JSPS and the Swiss National Science Foundation.

The authors thank Kazumi Toda-Peters for the experimental help in the use of the LightFab platform and Menouer Saidani and the nanofab support section of OIST for their technical support on the use of the Nanoscribe and the LightFab platform. The authors are also grateful to Shivani Sathish for the acquisition of the SEM images presented in Figure~\ref{fig:Chip}, and to Eliot Fried, Cameron Hopkins, Daniel Carlson and Stylianos Varchanis for the stimulating discussions related to the present project.

\bibliography{Main}

\end{document}